\documentclass[aps,prb,showpacs,twocolumn]{revtex4}

\usepackage[T1]{fontenc}
\usepackage{amssymb}
\usepackage{amsbsy}
\usepackage{amsmath}             
\usepackage{epsfig}

\begin{document}

\title{Effective action for superfluid Fermi systems in the strong-coupling
  limit} 
\author{N. Dupuis}
\affiliation{Department of Mathematics, Imperial College, 
180 Queen's Gate, London SW7 2AZ, UK \\ and \\ 
Laboratoire de Physique des Solides, CNRS UMR 8502, \\
  Universit\'e Paris-Sud, 91405 Orsay, France }

\date{December 20, 2004}


\begin{abstract}
We derive the low-energy effective action for three-dimensional superfluid
Fermi systems in the strong-coupling limit, where 
superfluidity originates from Bose-Einstein condensation of composite
bosons. Taking into account density and pairing fluctuations on the same
footing, we show that the effective action involves only the fermion density
$\rho_{\bf r}$ and its conjugate 
variable, the phase $\theta_{\bf r}$ of the pairing order parameter
$\Delta_{\bf r}$. We recover the standard
action of a Bose superfluid of density $\rho_{\bf r}/2$, where the bosons have
a mass 
$m_B=2m$ and interact {\it via} a repulsive contact potential with amplitude
$g_B=4\pi a_B/m_B$,  
$a_B=2a$ ($a$ the s-wave scattering length associated to the
fermion-fermion interaction in vacuum). For lattice models, the derivation of
the effective action is based on the mapping of the attractive Hubbard model
onto the Heisenberg model in a uniform magnetic field, and a coherent state
path integral representation of the partition function. The effective
description of the Fermi superfluid in the strong-coupling limit is a
Bose-Hubbard model with an intersite hopping amplitude $t_B=J/2$ and an
on-site repulsive interaction $U_B=2Jz$, where $J=4t^2/U$ ($t$ and $-U$ are 
the intersite hopping amplitude and the on-site attraction in the (fermionic)
Hubbard model, $z$ the number of nearest-neighbor sites). 
\end{abstract}

\pacs{74.20.Fg, 71.10.Fd, 05.30.Jp}

\maketitle

\section{Introduction}

Recent progress in the experimental control of ultracold atomic Fermi 
\cite{Kinast04,Bartenstein04,Zwierlein04,Regal04,Bourdel04}
gases has revived the interest in the crossover from the weak-coupling BCS
limit of superfluid fermions to the strong-coupling limit of condensed
composite bosons. \cite{Randeria95,BCS-BEC} In this paper, we derive the
low-energy effective action for a superfluid Fermi system in the
strong-coupling limit, both in continuum and lattice models. The latter may
be relevant for high-$T_c$ superconductors or ultracold Fermi gases in an
optical lattice.  

A Bose superfluid is described by a complex field $\psi_{B\bf r}=
\sqrt{\rho_{B\bf r}} e^{i\theta_{B\bf r}}$ where $\rho_{B\bf r}$ is the boson
density at position ${\bf r}$ in space. The equation of motion derived from the
standard action of a Bose system leads to the Gross-Pitaevskii
equation,\cite{Gross61,Pitaevskii61} i.e. a non-linear Schr\"odinger equation
for the $\psi_B$ field. The Gross-Pitaevskii equation yields a simple
description of quantum macroscopic phenomena like the Josephson effect or the
flux quantization,\cite{Feynman72,Ao95} and has proven to be a tool of choice
for the understanding of many phenomena in ultracold atomic Bose
gases. \cite{Pitaevskii03} In Fermi systems, there is in general
no simple relation between the amplitude of the superfluid (pairing) order
parameter $\Delta_{\bf r}$ and the fermion density $\rho_{\bf r}$. This
suggests that 
a minimal description, aiming at making contact with the standard description
of a Bose superfluid, should at least include the superfluid order parameter
$\Delta_{\bf r}$ and the density $\rho_{\bf r}$ from the outset. In the
strong-coupling 
limit, where superfluidity originates from Bose-Einstein condensation (BEC) of
composite bosons, we expect the description in terms of $\rho_{\bf r}$ and
$\Delta_{\bf r}=|\Delta_{\bf r}|e^{i\theta_{\bf r}}$ to be redundant and the
superfluid to be 
described by a single complex field $\psi_{\bf r}=\sqrt{\rho_{\bf r}/2}
e^{i\theta_{\bf r}}$ 
($\rho_{\bf r}/2$ being the density of composite bosons). 

Previous studies of the BCS-BEC crossover in superfluid Fermi systems can be
divided into two categories. In the first type of approach, 
\cite{Randeria95,Drechsler92,Haussmann93,deMelo93,Pistolesi96,Piery2000,Pieri03}
the density $\rho_{\bf r}$ is not 
considered explicitely and a pairing field $\Delta_{\bf r}^{\rm HS}$ is
introduced by means 
of a Hubbard-Stratonovich transformation of the fermion-fermion
interaction. In the BEC limit, the standard action $S[\psi^*,\psi]$ of a Bose
superfluid is recovered if one identifies $\psi_{\bf r}$ to $\Delta_{\bf
  r}^{\rm HS}$ (after 
a proper rescaling). For a continuum model, the bosons have a mass $m_B=2m$ and
interact {\it via} a repulsive contact potential with amplitude $g_B=4\pi
a_B/m_B$, $a_B=2a$ ($a$ is the s-wave scattering length associated to the
fermion-fermion interaction in vacuum). The main (conceptual) difficulty of
this approach 
is that the Hubbard-Stratonovich field $\Delta_{\bf r}^{\rm HS}$ is {\it not}
the physical 
pairing field $\Delta_{\bf r}=|\Delta_{\bf r}|e^{i\theta_{\bf r}}$ but rather
its conjugate 
field. \cite{Depalo99} Although both fields coincide at the mean-field level,
they differ when fluctuations are taken into account. As a result,
${\psi_{\bf r}}\propto\Delta_{\bf r}^{\rm HS}$ does not correspond to
$\sqrt{\rho_{\bf r}/2}e^{i\theta_{\bf r}}$ as expected.  

In the second type of approach to the BCS-BEC crossover,
\cite{Depalo99,Dupuis04,Stone95} the {\it physical} density and pairing
fields, $\rho_{\bf r}$ and $\Delta_{\bf r}$, are introduced from the
outset. For continuum models, only the weak-coupling limit has been
considered. \cite{Depalo99,Stone95}  
For lattice (Hubbard) models in the strong-coupling low-density limit, one
finds that the order parameter amplitude and the density are tied by the
relation $|\Delta_{\bf r}|=\sqrt{\rho_{\bf r}/2}$, so that the low-energy
effective action  
can be written in terms of a single complex field, $\Delta_{\bf r} =
\sqrt{\rho_{\bf r}/2} 
e^{i\theta_{\bf r}} \equiv \psi_{\bf r}$. \cite{Depalo99,Dupuis04} In the
continuum 
limit, one finds that the (composite) bosons have a mass $m_B=1/J$ and interact
{\it via} a repulsive contact potential with amplitude $g_B=8J$ (in two
dimensions), where $J=4t^2/U$ ($t$ being the intersite hopping amplitude and
$-U$ ($U\geq 0$) the on-site attractive interaction). \cite{Dupuis04}

Most of the theoretical works on the BCS-BEC crossover in ultracold atomic
Fermi gases have been formulated within a fermion-boson model,
\cite{fermion-boson} aiming at incorporating the molecular states involved in
the Feshbach resonance which drives the crossover. While the
equivalence of the fermion-boson 
model to an effective single-channel model in the crossover region may be
questionable, \cite{Falco04,Simonucci04} both models are equivalent in the
strong-coupling limit. 

The outline of the paper is as follows. In Sec.~\ref{sec:cm}, we extend the
approach of Ref.~\onlinecite{Depalo99} to the strong-coupling limit of a
continuum model. The particle-particle and particle-hole channels
are considered on the same footing, and the (physical) density
($\rho_{\bf r}$) and pairing ($\Delta_{\bf r}$) fields are 
introduced from the outset. The low-energy effective action is derived by
assuming small fluctuations of the collective fields about their mean-field
values. We find that fluctuations of $\rho_{\bf r}$ and $\Delta_{\bf r}$ are
not independent, 
so that the low-energy action can be written in term of a single complex field
${\psi_{\bf r}}=\sqrt{\rho_{\bf r}/2}e^{i\theta_{\bf r}}$. We recover the
standard action of a Bose 
superfluid with $m_B=2m$ and $g_B=4\pi a_B/m_B$, $a_B=2a$. For a lattice
model (Sec.~\ref{sec:lm}), we follow the approach introduced in
Ref.~\onlinecite{Dupuis04}. We map the attractive Hubbard model onto
the half-filled repulsive Hubbard model in a uniform magnetic field coupled to
the fermion spins. In the strong-coupling limit, the latter reduces to the
Heisenberg model in a uniform field. The low-energy effective action of the
attractive model is finally deduced from the coherent state path integral
representation of the Heisenberg model. In the low-density limit, where the
Pauli principle (which prevents two composite bosons to occupy the same site)
should not matter, $|\Delta_{\bf r}|\simeq \sqrt{\rho_{\bf r}/2}$ and the
superfluid Fermi 
system can be described by the complex field ${\psi_{\bf r}}= \sqrt{\rho_{\bf
    r}/2} 
e^{i\theta_{\bf r}}$. We find that the effective description of the Fermi
superfluid 
is a Bose-Hubbard model with intersite hopping amplitude $t_B=J/2$ and an
on-site repulsive interaction $U_B=2Jz$ (where $z$ is the number of
nearest-neighbor sites). 

\section{Continuum model}
\label{sec:cm} 

We consider a three-dimensional superfluid fermion system with the action
$S=S_0+S_{\rm int}$,  
\begin{eqnarray}
S_0 &=& {\int_0^\beta d\tau} {\int d^3{\bf r}\,}  c^\dagger_{\bf r}
\left(\partial_\tau - \mu - 
\frac{{\boldsymbol{\nabla}^2}}{2m} \right)  c_{\bf r} , \nonumber \\ 
S_{\rm int} &=& -g {\int_0^\beta d\tau} {\int d^3{\bf r}\,} c^*_{{\bf
    r}\uparrow}c^*_{{\bf r}\downarrow} 
c_{{\bf r}\downarrow} c_{{\bf r}\uparrow} ,
\label{action1} 
\end{eqnarray} 
where $c^{(*)}_{{\bf r}\sigma}$ are Grassmann variables, $ c_{\bf r}=(c_{{\bf
r}\uparrow},c_{{\bf r}\downarrow})^T$, $\tau$ an imaginary time and
$\beta=1/T$ the inverse temperature. $-g$ is the attractive interaction
between fermions ($g\geq 0$). The chemical potential $\mu$ fixes the average
fermion density $\rho_0$. To suppress ultraviolet divergences appearing in
the perturbation theory, one regularizes\cite{deMelo93} the local
fermion-fermion interaction 
with a cutoff $\Lambda$ acting on the fermion dispersion: $\epsilon_{\bf
  k}=|{\bf k}|^2/2m \leq \Lambda$.  $g$ and $\Lambda$ determine the s-wave
scattering length $a$ defined by the low-energy limit of the two-body problem
in vacuum,  
\begin{equation}
\frac{m}{4\pi a} = - \frac{1}{g} + \int_{\epsilon_{\bf k}\leq \Lambda}
\frac{d^3{\bf k}}{(2\pi)^3} \frac{1}{2\epsilon_{\bf k}} .
\label{scattering}
\end{equation}
$a$ is negative for small $g$ and diverges when $g=2\pi^2/m\Lambda$. For $g
>2\pi^2/m\Lambda$, there is a two-body bound-state (composite boson) with
energy $E_B=-1/ma^2$ and the scattering length $a$ is positive. The latter
also determines the extension of the bound-state. Low-energy
properties depend solely on $a$ (and not $g$ or $\Lambda$); we shall therefore
take the limit $g\to 0$ and $\Lambda\to\infty$ with $a$ fixed. In the
following, we consider the BEC limit defined by ${\rho_0 a^3}\ll 1$ ($a>0$),
where superfluidity originates from BEC of composite bosons. 

The (real) density and (complex) pairing fields,
\begin{eqnarray}
\rho_{\bf r} &=&  c^\dagger_{\bf r}  c_{\bf r} , \nonumber \\
S^z_{\bf r} &=&  c^\dagger_{\bf r} \sigma^z  c_{\bf r} , \nonumber \\ 
\Delta_{\bf r} &=&  c_{{\bf r}\downarrow} c_{{\bf r}\uparrow} , \nonumber \\
\Delta^*_{\bf r} &=& c^*_{{\bf r}\uparrow} c^*_{{\bf r}\downarrow} ,
\label{constraints}
\end{eqnarray}
can be introduced in the action by means of real ($ \rho^{\rm HS}_{\bf
  r},\tilde\rho^{\rm HS}_{\bf r}$) and 
complex ($\Delta_{\bf r}^{\rm HS}$) Lagrange multipliers:
\begin{eqnarray}
S &=& {\int_0^\beta d\tau} {\int d^3{\bf r}\,} \biggl\lbrace  c^\dagger_{\bf
  r} \biggl( \partial_\tau - \mu - 
\frac{{\boldsymbol{\nabla}^2}}{2m} \biggr)  c_{\bf r}  \nonumber \\ &&  
-g\gamma |\Delta_{\bf r}|^2 - \frac{g\alpha}{4}(\rho^2_{\bf r}-{S^z_{\bf r}}^2)
\nonumber \\ && 
+ i  \rho^{\rm HS}_{\bf r}(\rho_{\bf r}- c^\dagger_{\bf r} c_{\bf r})
+ i \tilde\rho^{\rm HS}_{\bf r} (S^z_{\bf r}- c^\dagger_{\bf r}\sigma^z c_{\bf
  r}) \nonumber \\ && 
+ i [\Delta_{\bf r}^{\rm HS}(\Delta^*_{\bf r}-c^*_{{\bf r}\uparrow}c^*_{{\bf
  r}\downarrow})  
+ {\rm c.c.}]  \biggr\rbrace .
\label{action2}
\end{eqnarray}
$(\sigma^x,\sigma^y,\sigma^z)$ denotes the Pauli matrices. 
Integrating over $ \rho^{\rm HS}_{\bf r},\tilde\rho^{\rm HS}_{\bf
  r},\Delta_{\bf r}^{\rm HS}$ and $\rho_{\bf r},S^z_{\bf 
  r},\Delta_{\bf r}$, we recover the original action (\ref{action1}) if we
choose $\alpha+\gamma=1$.\cite{note4} The
relative weights $\alpha$ and $\gamma$ of the particle-hole and
particle-particle channels are arbitrary. All the resulting effective actions
are equivalent when treated exactly. However, to recover the mean-field
results from a saddle-point approximation, we take $\alpha=\gamma=1$. When
only low-energy long-wavelength fluctuations about the mean-field state are
considered, there is no overlapping of the two channels and therefore no
overcounting. \cite{Depalo99} Note that by integrating out the 
physical fields $S^z_{\bf r}$, $\rho_{\bf r}$ and $\Delta_{\bf r}$, one
recovers the 
action $S[c, \rho^{\rm HS}_{\bf r},\tilde\rho^{\rm HS}_{\bf r},\Delta_{\bf
    r}^{\rm HS}]$ which is generally obtained by 
means of a Hubbard-Stratonovich decoupling of the interaction term. Thus the
Lagrange multipliers $ \rho^{\rm HS}_{\bf r}$, $\tilde\rho^{\rm HS}_{\bf r}$
and $\Delta_{\bf r}^{\rm HS}$ enforcing the 
constraints (\ref{constraints}) can also be seen as Hubbard-Stratonovich
fields.\cite{Depalo99}  In the following, we neglect spin fluctuations
($\tilde\rho^{\rm HS}_{\bf r}$ and $S^z_{\bf r}$) since they do not play an
important role when the interaction is attractive.

\subsection{Mean-field theory}

The mean-field theory is obtained from a saddle-point approximation where
the fields $\rho_{\bf r}$, $\Delta_{\bf r}$, $ \rho^{\rm HS}_{\bf r}$ and
$\Delta_{\bf r}^{\rm HS}$ are taken space and 
time independent. The saddle-point equations read
\begin{eqnarray}
\rho_0 = \langle  c^\dagger_{\bf r}  c_{\bf r} \rangle , && \,\,\,\, 
i\rho_0^{\rm HS} = \frac{g}{2} \rho_0 , \nonumber \\
\Delta_0 = \langle  c_{{\bf r}\downarrow} c_{{\bf r}\uparrow} \rangle , &&
\,\,\,\, 
i\Delta_0^{\rm HS} = g \Delta_0 , \nonumber \\ 
\Delta_0^* = \langle c^*_{{\bf r}\uparrow} c^*_{{\bf r}\downarrow} \rangle ,
&& \,\,\,\, 
i{\Delta_0^{\rm HS}}^* = g \Delta_0^* .
\label{speq}
\end{eqnarray} 
With no loss of generality, we can take $\Delta_0=\Delta_0^*$
real. $i\Delta_0^{\rm HS}=i{\Delta_0^{\rm HS}}^*$ is then real at the saddle
point. It is convenient to redefine 
$i\Delta_0^{\rm HS}\to \Delta_0^{\rm HS}$ and $i{\Delta_0^{\rm HS}}^*\to
{\Delta_0^{\rm HS}}^*$ (so that 
$\Delta_0^{\rm HS}={\Delta_0^{\rm HS}}^*$ is real) and absorb $i\rho_0^{\rm
  HS}$ in the definition of 
the chemical potential. The mean-field action is then (up to an additive
constant)
\begin{eqnarray}
S_{\rm MF} &=& {\int_0^\beta d\tau} {\int d^3{\bf r}\,} \biggl[
  c^\dagger_{\bf r} \biggl(\partial_\tau-\mu- 
  \frac{{\boldsymbol{\nabla}^2}}{2m} \biggr) c_{\bf r} \nonumber \\ && 
- \Delta_0^{\rm HS}(c^*_{{\bf r}\uparrow} c^*_{{\bf r}\downarrow}+{\rm c.c.})
  \biggr] . 
\label{actionhf}
\end{eqnarray}
From (\ref{actionhf}), we readily obtain the normal and anomalous Green
functions,
\begin{eqnarray}
G({\bf k},i\omega) &=& - \langle c_\sigma({\bf k},i\omega) c^*_\sigma({\bf
  k},i\omega) 
\rangle = \frac{-i\omega-{\xi_{\bf k}}}{\omega^2+{E^2_{\bf k}}} , \nonumber \\
F({\bf k},i\omega) &=&  - \langle c_\sigma({\bf k},i\omega)
  c_{\bar\sigma}(-{\bf k},-i\omega) 
\rangle = \frac{\Delta_0^{\rm HS}}{\omega^2+{E^2_{\bf k}}} , \nonumber \\
F^*({\bf k},i\omega) &=&  - \langle c^*_{\bar\sigma}(-{\bf k},-i\omega)
c^*_\sigma({\bf k},i\omega) \rangle = F({\bf k},i\omega) ,
\label{hfgf}
\end{eqnarray} 
where ${E_{\bf k}}=(\xi^2_{\bf k}+{\Delta_0^{\rm HS}}^2)^{1/2}$, ${\xi_{\bf
  k}}=\epsilon_{\bf 
  k}-\mu$, and 
$\bar\sigma=-\sigma$. $c_\sigma({\bf k},i\omega)$ is the Fourier transformed
field of $c_{{\bf r}\sigma}$ and $\omega$ a fermionic Matsubara frequency. 
Using (\ref{scattering}) and (\ref{hfgf}), we can rewrite the saddle-point
equations (\ref{speq}) as  
\begin{eqnarray}
\frac{m}{4\pi a} &=& {\int_{\bf k}} \biggl( \frac{1}{2\epsilon_{\bf k}} -
\frac{1}{2{E_{\bf k}}} 
\biggr) ,  \nonumber \\
\rho_0 &=& {\int_{\bf k}} \biggl( 1 - \frac{{\xi_{\bf k}}}{{E_{\bf k}}}
\biggr) , 
\label{speq1} 
\end{eqnarray}
where ${\int_{\bf k}}\equiv\int d^3{\bf k}/(2\pi)^3$. Eqs.~(\ref{speq1})
determine the 
chemical potential $\mu$ and the order parameter $\Delta_0^{\rm
  HS}=g\Delta_0$.  
In the strong-coupling limit ${\rho_0 a^3}\ll 1$, one obtains (see Appendix
\ref{app:hfcorr})  
\begin{eqnarray}
\mu &=& - \frac{1}{2ma^2}(1-2\pi {\rho_0 a^3}) , \nonumber \\
\Delta_0^{\rm HS} &=& \biggl(\frac{4\pi \rho_0}{m^2a}\biggr)^{1/2}
\biggl(1+\frac{\pi}{4}{\rho_0 a^3} \biggr) .  
\label{muhf} 
\end{eqnarray}

\subsection{Low-energy effective action}
\label{subsec:lea}

In this section, we derive the low-energy effective action for the physical
fields $\rho_{\bf r}$ and $\Delta_{\bf r}$. Since our derivation partially
follows 
Ref.~\onlinecite{Depalo99}, we describe only the main steps (technical details
are given in Appendix \ref{app:lea}).  
The main assumption is that the collective bosonic fields $\rho_{\bf r}$, $
\rho^{\rm HS}_{\bf r}$, 
$\Delta_{\bf r}$ and $\Delta_{\bf r}^{\rm HS}$ weakly fluctuate about their
mean-field values.  

Starting from the action (\ref{action2}) (with $\alpha=\gamma=1$), where 
\begin{equation}
\Delta_{\bf r} = |\Delta_{\bf r}| e^{i\theta_{\bf r}} ,
\end{equation}
we perform the change of variables 
\begin{equation}
 c_{\bf r} \to  c_{\bf r} e^{\frac{i}{2}\theta_{\bf r}} , \,\,\,\,
\Delta_{\bf r}^{\rm HS} \to \Delta_{\bf r}^{\rm HS} e^{i\theta_{\bf r}}  .
\end{equation}
We then consider the shift $ \rho^{\rm HS}_{\bf r} \to \rho_0^{\rm HS}+
\rho^{\rm HS}_{\bf r}$, $i\Delta_{\bf r}^{\rm HS} \to 
\Delta_0^{\rm HS}+i\Delta_{\bf r}^{\rm HS}$ and $i{\Delta_{\bf r}^{\rm HS}}^*
\to 
\Delta_0^{\rm HS}+i{\Delta_{\bf r}^{\rm HS}}^*$ (recall that 
a factor $i$ has been included in $\Delta_0^{\rm HS}$ and ${\Delta_0^{\rm
    HS}}^*$), so that the 
Hubbard-Stratonovich fields $ \rho^{\rm HS}_{\bf r}$ and $\Delta_{\bf r}^{\rm
  HS}$ now describe (small) 
fluctuations about the mean-field values. This leads to the action
\begin{eqnarray}
S &=& S_{\rm MF} + {\int_0^\beta d\tau} {\int d^3{\bf r}\,} \biggl[
  c^\dagger_{\bf r} \biggl( \frac{i}{2} 
  \dot\theta_{\bf r} - 
  \frac{i}{4m}{\boldsymbol{\nabla}\theta_{\bf r}} \cdot
  \tensor{\boldsymbol{\nabla}}  \nonumber \\ &&
+ \frac{({\boldsymbol{\nabla}\theta_{\bf r}})^2}{8m} 
  -\frac{\mu_B}{2} -i \rho^{\rm HS}_{\bf r} \biggr)  c_{\bf r}
- i (\Delta_{\bf r}^{\rm HS} c^*_{{\bf r}\uparrow}c^*_{{\bf r}\downarrow} +
  {\rm c.c.}) \nonumber \\ && 
+ (i{\Delta_{\bf r}^{\rm HS}}^*+i\Delta_{\bf r}^{\rm
  HS}+2\Delta_0^{\rm HS})|\Delta_{\bf r}| \nonumber \\ && 
-g |\Delta_{\bf r}|^2 - \frac{g}{4} \rho_{\bf r}^2 +(i\rho_0^{\rm HS}+i
  \rho^{\rm HS}_{\bf r})\rho_{\bf r} \biggr] , 
\end{eqnarray} 
where
$\tensor{\boldsymbol{\nabla}} = \roarrow{\boldsymbol{\nabla}} -
\loarrow{\boldsymbol{\nabla}}$.  
Here we write the chemical potential as $\mu=\mu_{\rm MF}+\mu_B/2$ where
$\mu_{\rm MF}$ is the chemical potential in the mean-field
approximation. The next step is to shift $ \rho^{\rm HS}_{\bf r}$, $i
\rho^{\rm HS}_{\bf r}\to i \rho^{\rm HS}_{\bf r}+ 
i\dot\theta_{\bf r}/2+({\boldsymbol{\nabla}\theta_{\bf r}})^2/8m-\mu_B/2$, and
to introduce Nambu spinors 
$\phi_{\bf r}=(c_{{\bf r}\uparrow},c^*_{{\bf r}\downarrow})^T$. This gives
\begin{eqnarray}
S &=& S_{\rm MF} + S' + {\int_0^\beta d\tau} {\int d^3{\bf r}\,} \biggl[
  -g|\Delta_{\bf r}|^2-\frac{g}{4} \rho_{\bf r}^2 
  \nonumber \\ && + (2\Delta_0^{\rm HS}+i\Delta_{\bf r}^{\rm HS}+i{\Delta_{\bf
      r}^{\rm HS}}^*)|\Delta_{\bf r}| \nonumber \\ 
  &&
+\rho_{\bf r} \Bigl(i\rho_0^{\rm HS}+i \rho^{\rm HS}_{\bf
  r}+\frac{i}{2}\dot\theta_{\bf r}+\frac{({\boldsymbol{\nabla}\theta_{\bf
  r}})^2}{8m} - 
\frac{\mu_B}{2}\Bigr) \biggr] , 
\end{eqnarray}
where 
\begin{eqnarray}
S' &=& {\int_0^\beta d\tau} {\int d^3{\bf r}\,} \Bigl(-i \rho^{\rm HS}_{\bf r}
{j^z_{0{\bf r}}} + \frac{1}{2} 
{\boldsymbol{\nabla}\theta_{\bf r}}  \cdot {\bf j}^0_{\bf r} 
\nonumber \\ && -
i\Delta_{\bf r}^{\rm HS} {j^+_{0{\bf r}}} - i {\Delta_{\bf r}^{\rm HS}}^*
{j^-_{0{\bf r}}} \Bigr)  ,  
\label{Sprime} \\ 
{j^z_{0{\bf r}}} &=& \phi^\dagger_{\bf r} \tau^z \phi_{\bf r} =
c^\dagger_{\bf r} 
c_{\bf r} , \nonumber \\  
{j^+_{0{\bf r}}} &=& \phi^\dagger_{\bf r} \tau^+ \phi_{\bf r} = c^*_{{\bf
    r}\uparrow} 
c^*_{{\bf r}\downarrow} , \nonumber 
\\  
{j^-_{0{\bf r}}} &=& \phi^\dagger_{\bf r} \tau^- \phi_{\bf r} =  c_{{\bf
    r}\downarrow} 
c_{{\bf r}\uparrow} ,  
\nonumber \\ 
{\bf j}^0_{\bf r} &=& -\frac{i}{2m} \phi^\dagger_{\bf r} \tensor
{\boldsymbol{\nabla}} 
\phi_{\bf r} = -\frac{i}{2m}  c^\dagger_{\bf r} \tensor {\boldsymbol{\nabla}}
c_{\bf r} . 
\label{jdef}  
\end{eqnarray} 
$(\tau_x,\tau_y,\tau_z)$ are Pauli
matrices acting in Nambu space. The effective action
$S[\rho, \rho^{\rm HS},\Delta,\Delta^{\rm HS}]$ is obtained by integrating out
the 
fermions. To quadratic order in the bosonic fields and their gradient
($\partial_\tau$ or $\boldsymbol{\nabla}$),  it is sufficient to retain the
first and second order cumulants of $S'$ with respect to the mean-field
action:   
\begin{multline}
S[\rho, \rho^{\rm HS},\Delta,\Delta^{\rm HS}] = \biggl\langle
  S'-\frac{{S'}^2}{2} 
\biggr\rangle_c \\
+ {\int_0^\beta d\tau} {\int d^3{\bf r}\,} \biggl[ -g|\Delta_{\bf r}|^2 -
  \frac{g}{4} \rho_{\bf r}^2  \\ 
+ (2\Delta_0^{\rm HS}+i{\Delta_{\bf r}^{\rm HS}}^*+i\Delta_{\bf r}^{\rm
  HS})|\Delta_{\bf r}| \\ 
+\rho_{\bf r} \Bigl(i\rho_0^{\rm HS}+i \rho^{\rm HS}_{\bf
  r}+\frac{i}{2}\dot\theta_{\bf r}+\frac{({\boldsymbol{\nabla}\theta_{\bf
  r}})^2}{8m} - 
\frac{\mu_B}{2}\Bigr) \biggr] ,
\label{action_cf}
\end{multline}
where the averages $\langle \cdots\rangle_c$ are calculated with respect to 
the mean-field action $S_{\rm MF}$. 
Calculating the first and second order cumulants and integrating out the
Hubbard-Stratonovich fields $ \rho^{\rm HS}_{\bf r}$ and $\Delta_{\bf r}^{\rm
  HS}$ (Appendix \ref{app:lea}), we obtain  
\begin{multline} 
S[\rho,\Delta] = {\int_0^\beta d\tau} {\int d^3{\bf r}\,} \rho_{\bf r}
\biggl(\frac{i}{2}\dot\theta_{\bf r}+\frac{({\boldsymbol{\nabla}\theta_{\bf
      r}})^2}{8m}  
-\frac{\mu_B}{2} \biggr) \\
+ \sum_q (\delta\rho_{-q},\delta|\Delta|_{-q})
\left(
\begin{array}{ll}
{\beta_{\bf q}} & -g{\alpha_{\bf q}} \\
-g{\alpha_{\bf q}} & g^2 {\gamma_{\bf q}} 
\end{array}
\right) \left(
\begin{array}{l}
\delta\rho_q \\ \delta|\Delta|_q
\end{array}
\right) , 
\label{action3} 
\end{multline}
where $\delta\rho_q$ and $\delta|\Delta|_q$ are the Fourier transforms of
$\delta\rho_{\bf r}=\rho_{\bf r}-\rho_0$ and $\delta|\Delta_{\bf
  r}|=|\Delta_{\bf r}|-\Delta_0$, and  
\begin{eqnarray}
{\beta_{\bf q}} &=&  \frac{1}{2}{\Pi^{zz}_{00}({\bf q})}^{-1}-\frac{g}{4} -
    \frac{1}{{C_{\bf q}}} [{\Pi^{zz}_{00}({\bf q})}^{-1}{\Pi^{z+}_{00}({\bf
    q})}]^2 , \nonumber \\ 
-g {\alpha_{\bf q}} &=& 
\frac{1}{C_q} {\Pi^{zz}_{00}({\bf q})}^{-1} {\Pi^{z+}_{00}({\bf q})}  ,
\nonumber \\ 
g^2 {\gamma_{\bf q}} &=& -\frac{1}{{C_{\bf q}}}-g , \nonumber \\ 
{C_{\bf q}} &=& - {\Pi^{+-}_{00}({\bf q})} - {\Pi^{++}_{00}({\bf q})} +
    2{\Pi^{zz}_{00}({\bf q})}^{-1} {\Pi^{z+}_{00}({\bf q})}^2 . 
\nonumber \\ && 
\label{action3a}
\end{eqnarray} 
We use the notation $q=({\bf q},i\omega_\nu)$ and $\sum_q=\sum_{{\bf
q},\omega_\nu}$ where $\omega_\nu$ is a bosonic Matsubara frequency. 
The mean-field correlation function $\Pi^{\nu\nu'}_{00}(q)= \langle
j^\nu_0(q) j^{\nu'}_0(-q)\rangle_c$ is calculated in Appendix
\ref{app:hfcorr} and $\Pi^{\nu\nu'}_{00}({\bf q})=\Pi^{\nu\nu'}_{00}({\bf
  q},\omega_\nu=0)$. $j^\nu_0(q)$ is the Fourier transformed field of
$j^\nu_{0{\bf r}}$ [Eq.~(\ref{jdef})]. Eq.~(\ref{action3}) shows that half the
fermion density is the conjugate variable of the phase $\theta_{\bf r}$ of the
pairing field. Eqs.~(\ref{action3}-\ref{action3a}) agree with Eq.~(2.3) of 
Ref.~\onlinecite{Depalo99} except for the coefficient of $\delta\rho_{-q}
\delta|\Delta|_q$ which is found to have opposite sign.\cite{note5} 

We now discuss the strong-coupling limit (not considered in
Ref.~\onlinecite{Depalo99}). To leading order in ${\rho_0 a^3}$
and $|{\bf q}|a$, we have (Appendix \ref{app:hfcorr}) \cite{Marini98} 
\begin{eqnarray}
{\alpha_{\bf q}} &=& \biggl(\frac{1}{4\pi{\rho_0 a^3}}\biggr)^{1/2} \biggl( 1 +
\frac{9}{4}\pi{\rho_0 a^3}+\frac{1}{6}|{\bf q}|^2a^2 \biggr) , \nonumber \\
{\beta_{\bf q}} &=& \frac{1}{2\rho_0 ma^2} \biggl(1+4\pi{\rho_0
  a^3}+\frac{1}{4}|{\bf q}|^2a^2 
\biggr) , \nonumber \\ 
{\gamma_{\bf q}} &=& \frac{m}{2\pi a}  \biggl( 1 + \frac{3}{2}\pi{\rho_0
  a^3}+\frac{7}{48} 
|{\bf q}|^2a^2 \biggr)  . 
\label{abg} 
\end{eqnarray} 
Denoting by $\lambda^+_{\bf q}$ and  $\lambda^-_{\bf q}$  the two eigenvalues
of the fluctuation matrix appearing in (\ref{action3}), we have 
\begin{eqnarray}
\lambda^+_{\bf q} &=& {\beta_{\bf q}} + g^2 \frac{{\alpha_{\bf
      q}}^2}{{\beta_{\bf q}}} , \nonumber \\  
\lambda^-_{\bf q} &=& g^2 \biggl( {\gamma_{\bf q}} - \frac{{\alpha_{\bf
      q}}^2}{{\beta_{\bf q}}} \biggr) ,
\end{eqnarray}
to order ${\cal O}(g^2)$. For $g\to 0$ (at fixed $a$), the mode corresponding
to the eigenvalue $\lambda^+_{\bf q}$ is frozen, which leads to 
\begin{equation}
\frac{\delta|\Delta_{\bf r}|}{\delta\rho_{\bf r}} = \frac{1}{g} \biggl(
  \frac{\pi}{\rho_0 m^2 
  a} \biggr)^{1/2}.
\label{tied} 
\end{equation}
Density ($\delta\rho_{\bf r}$) and modulus ($\delta|\Delta_{\bf r}|$)
fluctuations do not 
fluctuate independently in the low-energy limit but are tied by the relation
(\ref{tied}). From (\ref{action3},\ref{abg},\ref{tied}), we deduce that 
the dynamics of the Fermi superfluid is determined by the effective action
\begin{eqnarray}
S[\rho,\theta] &=& {\int_0^\beta d\tau} {\int d^3{\bf r}\,} \biggl[ \rho_{\bf
  r} \biggl( 
  \frac{i}{2}\dot\theta_{\bf r} + 
  \frac{({\boldsymbol{\nabla}\theta_{\bf r}})^2}{8m} - \frac{\mu_B}{2} \biggr)
  \nonumber \\  
&& + \frac{\pi a}{2m} (\delta\rho_{\bf r})^2 +
  \frac{(\boldsymbol{\nabla}\delta\rho_{\bf 
  r})^2}{32\rho_0 m} \biggr] .
\label{action4}
\end{eqnarray} 
Introducing the bosonic field
\begin{equation}
{\psi_{\bf r}} = \sqrt{\frac{\rho_{\bf r}}{2}} e^{i\theta_{\bf r}} ,
\label{psidef}
\end{equation}
we recover the standard action of a Bose superfluid,
\begin{eqnarray}
S[\psi^*,\psi] &=& {\int_0^\beta d\tau} {\int d^3{\bf r}\,} \biggl[
  {\psi^*_{\bf r}} \biggl(\partial_\tau - \mu_B - 
  \frac{{\boldsymbol{\nabla}^2}}{2m_B} \biggr) {\psi_{\bf r}} \nonumber \\ 
&& + \frac{2\pi a_B}{m_B} ({\psi^*_{\bf r}}{\psi_{\bf r}}-\rho_0/2)^2 \biggr], 
\label{action5}
\end{eqnarray}
where $m_B = 2m$ and $a_B = 2a$ are the mass and the scattering length of the
bosons. The result $a_B=2a$ corresponds to the Born approximation for the
boson-boson scattering, while the exact result is $a_B=0.6a$. \cite{Petrov04}
Eqs.~(\ref{action5}) and  
(\ref{action4}) are equivalent in the hydrodynamic regime where
$(\boldsymbol{\nabla} 
\rho_{\bf r})^2/\rho_0\simeq (\boldsymbol{\nabla}\rho_{\bf r})^2/\rho_{\bf
  r}$. \cite{note2}  

Thus, we have shown how, by introducing the physical fields $\rho_{\bf r}$ and
$\Delta_{\bf r}$ from the outset and expanding about the mean-field state in
the 
strong-coupling limit, one obtains the standard action of a Bose
superfluid. Our approach should be contrasted with a number of previous works 
\cite{Randeria95,Drechsler92,Haussmann93,deMelo93,Pistolesi96,Piery2000,Pieri03}  
where only the pairing Hubbard-Stratonovich field $\Delta_{\bf r}^{\rm HS}$ is
considered and the expansion is carried out about the non-interacting
state, which gives the action (\ref{action5}) but for the field
$\sqrt{\rho_{\bf r}/2}e^{i\theta^{\rm HS}_{\bf r}}$ instead of the $\psi$
field defined in (\ref{psidef}). \cite{note3}

\section{Lattice model}
\label{sec:lm}

In this section, we consider the attractive Hubbard model on a bipartite
lattice, with Hamiltonian 
\begin{equation}
H = -t \sum_{\langle{\bf r},{\bf r}'\rangle}  (c^\dagger_{\bf r} c_{{\bf r}'}
+{\rm h.c.}) 
- \mu {\sum_{\bf r}}  c^\dagger_{\bf r}  c_{\bf r} - U {\sum_{\bf r}} {n_{{\bf
      r}\uparrow}} {n_{{\bf r}\downarrow}} . 
\label{ham1}
\end{equation}
The operator $c^\dagger_{{\bf r}\sigma}$ ($c_{{\bf r}\sigma}$) creates
(annihilates) a fermion with spin $\sigma$ at the lattice site ${\bf r}$,
$ c_{\bf r}=(c_{{\bf r}\uparrow},c_{{\bf r}\downarrow})^T$, and $n_{{\bf
    r}\sigma}=c^\dagger_{{\bf r}\sigma}c_{{\bf r}\sigma}$. $\langle{\bf
  r},{\bf r}'\rangle$ denotes nearest-neighbor sites. The chemical potential
$\mu$ fixes the average density $\rho_0$ (i.e. the average number of fermions
per site) and $-U$ ($U\geq 0$) is the on-site attractive interaction. 

We are interested in the strong-coupling limit $U\gg t$ where fermions form
tightly bound composite bosons which behave as local pairs. The latter Bose
condense at low temperature giving rise to superfluidity. In order to derive
the low-energy effective action, we could follow the procedure used in 
Sec.~\ref{sec:cm}. Here, we shall use a different method, based on the
mapping of the attractive Hubbard model in the strong-coupling limit onto the
Heisenberg model in a uniform magnetic field. \cite{Dupuis04} Thus this
approach is based on a $t/U$ expansion about the $t=0$ limit rather than
on an expansion about the mean-field state. \cite{note1}

Under the canonical particle-hole transformation \cite{Micnas90}
\begin{equation}
c_{{\bf r}\downarrow} \to (-1)^{\bf r} c^\dagger_{{\bf r}\downarrow} , \,\,\, 
c^\dagger_{{\bf r}\downarrow} \to (-1)^{\bf r} c_{{\bf r}\downarrow} ,
\end{equation}
the Hamiltonian becomes (omitting a constant term)
\begin{eqnarray}
H &=& -t \sum_{\langle{\bf r},{\bf r}'\rangle} ( c^\dagger_{\bf r} c_{{\bf
    r}'}  +{\rm h.c.})
- {\sum_{\bf r}}  c^\dagger_{\bf r} \biggl(\frac{U}{2} + h_0 \sigma^z \biggr)
c_{\bf r} \nonumber \\ && 
+ U {\sum_{\bf r}} {n_{{\bf r}\uparrow}} {n_{{\bf r}\downarrow}} ,
\label{ham2}
\end{eqnarray}
and corresponds now to the repulsive Hubbard model in a magnetic
field ${\bf h}_0=h_0 \hat {\bf z}$ along the $z$ axis, 
\begin{equation}
h_0 = \mu + \frac{U}{2},
\end{equation}
coupled to the fermion spins. The chemical potential $U/2$ in (\ref{ham2}),
together with particle-hole symmetry, implies that the system is half-filled. 
The density and pairing operators transform into the three components of the
spin density operator: 
\begin{eqnarray}
\rho_{\bf r} =  c^\dagger_{\bf r}  c_{\bf r} &\to&  c^\dagger_{\bf r} \sigma^z
c_{\bf r} +1 , \nonumber \\   
\Delta_{\bf r} =  c_{{\bf r}\downarrow} c_{{\bf r}\uparrow} &\to& (-1)^{\bf r}
c^\dagger_{{\bf r}\downarrow} c_{{\bf r}\uparrow} ,  
\nonumber \\  
\Delta^\dagger_{\bf r} = c^\dagger_{{\bf r}\uparrow} c^\dagger_{{\bf r}
  \downarrow} &\to& (-1)^{\bf r} c^\dagger_{{\bf r}\uparrow}  c_{{\bf
    r}\downarrow}  . 
\label{correspondence}
\end{eqnarray}
The equation fixing $\mu$, $\langle  c^\dagger_{\bf r} c_{\bf r}
\rangle=\rho_0$, becomes an equation 
fixing the magnetic field: $\langle  c^\dagger_{\bf r} \sigma^z c_{\bf
  r}\rangle=\rho_0-1$. 

In the strong-coupling limit $U\gg t$, the Hamiltonian (\ref{ham2}) simplifies
into \cite{Dupuis04}
\begin{equation}
H = J \sum_{\langle {\bf r},{\bf r}'\rangle} {\bf S_r}\cdot {\bf S}_{{\bf r}'}
- 2 {\bf h}_0 \cdot {\sum_{\bf r}} {\bf S_r} ,
\label{ham3}
\end{equation}
where $J=4t^2/U$ and ${\bf S_r}$ is a spin-$\frac{1}{2}$ operator. Using
spin-$\frac{1}{2}$ coherent states $|{\bf\Omega_r}\rangle$ (${\bf\Omega}^2_{\bf
  r}=1$), \cite{Auerbach94} the action of the Heisenberg model (\ref{ham3})
can be written as 
\begin{equation}
S[{\bf\Omega}] = {\int_0^\beta d\tau} \biggl[ {\sum_{\bf r}}
  [\langle{\bf\Omega_r}|\dot{\bf\Omega}_{\bf r} 
  \rangle - {\bf h}_0 \cdot {\bf\Omega_r}] + J \sum_{\langle{\bf r},{\bf
   r}'\rangle} \frac{{\bf\Omega_r} \cdot {\bf\Omega}_{{\bf r}'}}{4} \biggr] ,
\label{action_heis}
\end{equation} 
where $|\dot{\bf\Omega}_{\bf r}\rangle=\partial_\tau|{\bf\Omega_r}\rangle$. 

The effective action $S[\rho,\Delta]$ of the superfluid system is obtained by
rewriting the action (\ref{action_heis}) in terms of the density and pairing
fields of the attractive model. In the strong-coupling limit,
Eqs.~(\ref{correspondence}) (written now for fields rather than operators)
become \cite{Dupuis04}  
\begin{eqnarray}
\rho_{\bf r} &=& \Omega^z_{\bf r} + 1 , \nonumber \\ 
\Delta_{\bf r} &=& \frac{(-1)^{\bf r}}{2} \Omega^-_{\bf r} , \nonumber \\
\Delta^*_{\bf r} &=& \frac{(-1)^{\bf r}}{2} \Omega^+_{\bf r} ,
\label{correspondence2}
\end{eqnarray}
where $\Omega^\pm_{\bf r}=\Omega^x_{\bf r}\pm i\Omega^y_{\bf r}$. The
condition ${\bf\Omega}^2_{\bf r}=1$ implies that $\rho_{\bf r}$ and
$\Delta_{\bf r}$ do not 
fluctuate independently but are tied by the relation
\begin{equation}   
|\Delta_{\bf r}| = \frac{1}{2} [\rho_{\bf r}(2-\rho_{\bf r})]^{1/2} .
\label{tied1}
\end{equation}  

In the low-density limit ($\rho_{\bf r}\ll 1$), where the Pauli principle
(which 
prevents two composite bosons to occupy the same site) should not matter,
we expect to recover the standard action of a Bose superfluid. In that limit,
$|\Delta_{\bf r}|\simeq\sqrt{\rho_{\bf r}/2}$; the pair density $|\Delta_{\bf
  r}|^2$ equals half 
the fermion density $\rho_{\bf r}$, and $\Delta_{\bf r}=|\Delta_{\bf
  r}|e^{i\theta_{\bf r}}$ coincides   
with the bosonic field ${\psi_{\bf r}}=\sqrt{\rho_{\bf r}/2}e^{i\theta_{\bf
    r}}$.  
To order ${\cal O}(\rho^2_{\bf r})$, we deduce from
(\ref{action_heis}-\ref{tied1}) 
\begin{eqnarray}
S[\rho,\theta] &=& {\int_0^\beta d\tau} \biggl\lbrace  
{\sum_{\bf r}} \biggl[ \frac{i}{2} \rho_{\bf r}\dot\theta_{\bf r} -
  \biggl(h_0+\frac{Jz}{4}\biggr)\rho_{\bf r}  \biggr] \nonumber \\ && 
+ \frac{J}{4} \sum_{\langle {\bf r},{\bf r}'\rangle} [\rho_{\bf r}\rho_{{\bf
      r}'}  
\nonumber \\ && 
- (\rho_{\bf r}\rho_{{\bf r}'})^{1/2}(2-\rho_{\bf r})\cos(\theta_{\bf
  r}-\theta_{{\bf r}'}) ]  
\biggr\rbrace .
\end{eqnarray}
The term
$(i/2)\rho_{\bf r}\dot\theta_{\bf r}$ comes from the Berry phase term
$\langle{\bf\Omega_r}|\dot{\bf\Omega}_{\bf r} \rangle$ of the action
$S[{\bf\Omega}]$ [Eq.~(\ref{action_heis})] with a proper gauge
choice. \cite{Dupuis04} If we further assume that $\rho_{\bf r}$ and
$\theta_{\bf r}$ are slowly varying in space, we obtain
\begin{eqnarray}
S[\rho,\theta] &=& {\int_0^\beta d\tau} \biggl\lbrace 
{\sum_{\bf r}} \biggl[ \frac{i}{2} \rho_{\bf r}\dot\theta_{\bf r} -
  \biggl(h_0+\frac{Jz}{4} \biggr) \rho_{\bf r} 
+ \frac{Jz}{4} \rho^2_{\bf r} \biggr] \nonumber \\ &&
- \frac{J}{2} \sum_{\langle {\bf r},{\bf r}'\rangle} (\rho_{\bf r}\rho_{{\bf
    r}'})^{1/2} \cos(\theta_{\bf r}-\theta_{{\bf r}'}) \biggr\rbrace .
\end{eqnarray}
or, equivalently,
\begin{eqnarray}
S[\psi^*,\psi] &=& {\int_0^\beta d\tau} \biggl\lbrace
{\sum_{\bf r}}  \biggl[{\psi^*_{\bf r}}(\partial_\tau - \mu_B) {\psi_{\bf r}}
  + \frac{U_B}{2} 
  |\psi^4_{\bf r}| \biggr] \nonumber \\ &&
- t_B \sum_{\langle {\bf r},{\bf r}'\rangle} ({\psi^*_{\bf r}}  \psi_{{\bf
    r}'} + {\rm c.c.}) \biggr\rbrace ,
\end{eqnarray}
where $t_B=J/2$, $U_B=2Jz$, $\mu_B=2h_0+Jz/2$, and $z$ is the number of
nearest-neighbor sites. We therefore obtain the
action of the Bose-Hubbard model with on-site repulsive interaction $U_B$ and
nearest-neighbor hopping amplitude $t_B$. In the continuum limit and for a
cubic lattice, the latter gives a boson mass $m_B=1/J$ as obtained in
Ref.~\onlinecite{Dupuis04}.

\section{Concluding remarks}

In this paper, we have shown that a Fermi superfluid in the strong-coupling
limit, where superfluidity originates from BEC of composite bosons, can be
described by the complex field $\psi_{\bf r}=\sqrt{\rho_{\bf
    r}/2}e^{i\theta_{\bf r}}$, 
where $\rho_{\bf r}$ is the fermion density and $\theta_{\bf r}$ the phase of
the pairing 
field $\Delta_{\bf r}$. Such is description is made possible by the fact that
density 
($\rho_{\bf r}$) and amplitude ($|\Delta_{\bf r}|$) fluctuations are not
independent in the 
strong-coupling limit. The effective action $S[\rho,\theta]$ is derived by
introducing the physical fields $\rho_{\bf r}$ and $\Delta_{\bf r}$ from the
outset by means 
of Lagrange multiplier fields $ \rho^{\rm HS}_{\bf r}$ and $\Delta_{\bf
  r}^{\rm HS}$. The latter play the 
role of the Hubbard-Stratonovich fields usually introduced {\it via} a
Hubbard-Stratonovich transformation of the fermion-fermion interaction. 

For continuum models, the effective action is derived from an expansion about
the mean-field state. It corresponds to the usual action of a Bose superfluid
of density $\rho_{\bf r}/2$ where the bosons have a mass $m_B=2m$ and interact
{\it via} a contact potential with amplitude $g_B=4\pi a_B/m_B$, $a_B=2a$. 

For lattice (Hubbard) models, the effective action is obtained from an
expansion about the $t=0$ limit, using the mapping of the attractive Hubbard
model in the strong-coupling limit onto the Heisenberg model in a uniform
magnetic field. The effective model is a Bose-Hubbard model with an on-site
repulsion $U_B=2Jz$ (with $z$ the number of nearest-neighbor sites) and a
nearest-neighbor intersite hopping amplitude $t_B=J/2$, where $J=4t^2/U$.


\appendix

\begin{widetext} 

\section{Low-energy effective action $S[\rho,\Delta]$}
\label{app:lea}

In this appendix, we derive the effective action (\ref{action3}) starting from
Eq.~(\ref{action_cf}). The first and second order cumulants are given by
\begin{eqnarray}
\langle S' \rangle &=& {\int_0^\beta d\tau} {\int d^3{\bf r}\,} [-i\rho_0
  \rho^{\rm HS}_{\bf r} -i \Delta_0 
  (\Delta_{\bf r}^{\rm HS}+{\Delta_{\bf r}^{\rm HS}}^*) ] , \nonumber \\ 
\langle {S'}^2 \rangle_c &=& \sum_q \biggl[ -\rho_{-q}^{\rm HS}
  {\Pi^{zz}_{00}(q)} \rho_{q}^{\rm HS} - 2 \rho_{-q}^{\rm HS}
  {\Pi^{z+}_{00}(q)} {\Delta_{q}^{\rm HS}}  
- 2 \rho_{-q}^{\rm HS} {\Pi^{z-}_{00}(q)} {{\Delta_{-q}^{\rm HS}}^*} 
- {\Delta_{-q}^{\rm HS}} {\Pi^{++}_{00}(q)}  {\Delta_{q}^{\rm HS}} \nonumber
  \\ && 
 - {{\Delta_{q}^{\rm HS}}^*} {\Pi^{--}_{00}(q)} {{\Delta_{-q}^{\rm HS}}^*} - 2
  {\Delta_{-q}^{\rm HS}} 
{\Pi^{+-}_{00}(q)} {{\Delta_{-q}^{\rm HS}}^*} \biggr] .
\label{cumulant}
\end{eqnarray} 
The second order cumulant is written in Fourier space. The mean-field
correlation function $\Pi^{\nu\nu'}_{00}(q) = \langle j^\nu_0(q)
j^{\nu'}_0(-q) \rangle_c$ is defined in Sec.~\ref{subsec:lea}. To obtain
(\ref{cumulant}), 
we have used the fact that mean-field correlation functions involving the
current ${\bf j}^0_{\bf r}$ vanish (Appendix \ref{app:hfcorr}). In the
low-energy limit, we 
can approximate $\Pi^{\nu\nu'}_{00}(q)$ by its static limit
$\Pi^{\nu\nu'}_{00}({\bf q})=\Pi^{\nu\nu'}_{00}({\bf q},\omega_\nu=0)$. 

Integrating out the  Hubbard-Stratonovich field $ \rho^{\rm HS}_{\bf r}$, we
obtain 
\begin{eqnarray}
S &=& {\int_0^\beta d\tau} {\int d^3{\bf r}\,} \biggl[ - g\delta|\Delta_{\bf
  r}|^2 - 
  \frac{g}{4} \delta\rho_{\bf r}^2
+ \rho_{\bf r} \Bigl( \frac{i}{2} \dot\theta_{\bf r} +
  \frac{({\boldsymbol{\nabla}\theta_{\bf r}})^2}{8m}  
- \frac{\mu_B}{2} \Bigr) 
+i ({\Delta_{\bf r}^{\rm HS}}^* +\Delta_{\bf r}^{\rm HS})\delta|\Delta_{\bf
  r}| \biggr] \nonumber  \\ && 
+ \sum_q \biggl\lbrace \frac{1}{2} {\Pi^{++}_{00}({\bf q})} ({\Delta_{-q}^{\rm
  HS}} {\Delta_{q}^{\rm HS}} + 
{\rm c.c.}) +{{\Delta_{q}^{\rm HS}}^*} {\Pi^{+-}_{00}({\bf q})}
  {\Delta_{q}^{\rm HS}} \nonumber \\ && 
+ \frac{1}{2}{\Pi^{zz}_{00}({\bf q})}^{-1} \Bigl[ \delta\rho_{-q} \delta\rho_q
  - 2i {\Pi^{z+}_{00}(q)} 
  \delta\rho_{-q}({\Delta_{q}^{\rm HS}}+{{\Delta_{-q}^{\rm HS}}^*})
- {\Pi^{z+}_{00}({\bf q})}^2({\Delta_{-q}^{\rm HS}}{\Delta_{q}^{\rm
  HS}}+{{\Delta_{-q}^{\rm HS}}^*} {\Delta_{q}^{\rm HS}} + 2 
|{\Delta_{q}^{\rm HS}}|^2) \Bigr] \biggr\rbrace ,
\label{appA1}
\end{eqnarray} 
where $\delta\rho_{\bf r} = \rho_{\bf r}-\rho_0$ and $\delta|\Delta_{\bf
  r}|=\Delta_{\bf r}-\Delta_0$. Here
we have neglected constant terms and use the saddle-point equations
(\ref{speq}).  

To obtain the action $S[\rho,\Delta]$ in terms of the physical fields only,
one has then to integrate out the Hubbard-Stratonovich field $\Delta^{\rm HS}$:
\begin{eqnarray}
&& \int {\cal D}[\Delta^{\rm HS}] \exp\Bigl\lbrace-S_0[\Delta^{\rm HS}] -i
  \sum_q 
  ({\Delta_{q}^{\rm HS}}+{{\Delta_{-q}^{\rm HS}}^*})
  [\delta|\Delta|_{-q}-{\Pi^{zz}_{00}({\bf q})}^{-1}{\Pi^{z+}_{00}({\bf q})} 
    \delta\rho_{-q}]\Bigr\rbrace \nonumber \\ 
&=& 
\exp \Bigl\lbrace -\frac{1}{2} \sum_q \langle ({\Delta_{q}^{\rm
  HS}}+{{\Delta_{-q}^{\rm 
  HS}}^*}) 
({\Delta_{-q}^{\rm HS}}+{{\Delta_{q}^{\rm HS}}^*}) \rangle_0
[\delta|\Delta|_{q}-{\Pi^{zz}_{00}({\bf q})}^{-1}{\Pi^{z+}_{00}({\bf q})}
  \delta\rho_q] 
[\delta|\Delta|_{-q}-{\Pi^{zz}_{00}({\bf q})}^{-1}{\Pi^{z+}_{00}({\bf q})}
  \delta\rho_{-q}] 
\Bigr\rbrace \nonumber \\ 
&=&
\exp \Bigl\lbrace \frac{1}{2} \sum_q [M_{11}({\bf q})+M_{22}({\bf
  q})+2M_{12}({\bf q})] [\delta|\Delta|_{q}-{\Pi^{zz}_{00}({\bf
  q})}^{-1}{\Pi^{z+}_{00}({\bf q})} \delta\rho_q] 
[\delta|\Delta|_{-q}-{\Pi^{zz}_{00}({\bf q})}^{-1}{\Pi^{z+}_{00}({\bf q})}
  \delta\rho_{-q}] \Bigr\rbrace, 
\label{appA2}
\end{eqnarray}
where averages $\langle \cdots \rangle_0$ are taken with the Gaussian action 
\begin{eqnarray}
S_0[\Delta^{\rm HS}] &=& - \frac{1}{2} \sum_q ({{\Delta_{q}^{\rm
  HS}}^*},{\Delta_{-q}^{\rm HS}}) 
  M^{-1}({\bf q})  
\left(
\begin{array}{l}
{\Delta_{q}^{\rm HS}}  \\ {{\Delta_{-q}^{\rm HS}}^*}  
\end{array}
\right) , \nonumber \\ 
M^{-1}({\bf q}) &=& \left(
\begin{array}{lr}
-{\Pi^{+-}_{00}({\bf q})} + {\Pi^{zz}_{00}({\bf q})}^{-1} {\Pi^{z+}_{00}({\bf
    q})}^2 & -{\Pi^{++}_{00}({\bf q})}+{\Pi^{zz}_{00}({\bf
    q})}^{-1}{\Pi^{z+}_{00}({\bf q})}^2 \\ 
 -{\Pi^{++}_{00}({\bf q})}+{\Pi^{zz}_{00}({\bf q})}^{-1}{\Pi^{z+}_{00}({\bf
    q})}^2 & -{\Pi^{+-}_{00}({\bf q})} + {\Pi^{zz}_{00}({\bf q})}^{-1}
    {\Pi^{z+}_{00}({\bf q})}^2  
\end{array}
\right) .
\end{eqnarray} 
In the following, we denote by ${A_{\bf q}}$ and ${B_{\bf q}}$ the diagonal and
off-diagonal 
components of $M^{-1}({\bf q})$, and ${C_{\bf q}}={A_{\bf q}}+{B_{\bf q}}$. The
effective action 
$S[\rho,\Delta]$ deduced from Eqs.~(\ref{appA1},\ref{appA2}) is given by
Eq.~(\ref{action3}).  
 
\end{widetext}

\section{Mean-field correlation function} 
\label{app:hfcorr}

In this appendix, we calculate the mean-field correlation function
$\Pi^{\nu\nu'}_{\mu\mu'}(q) = \langle j^\nu_\mu(q) j^{\nu'}_{\mu'}(-q)
\rangle_c$ ($\nu,\nu'=x,y,z$; $\mu,\mu'=0,x,y,z$) in the strong-coupling limit
${\rho_0 a^3}\ll 1$ and for $|{\bf q}|a\ll 1$. 

\subsection{General expression}

$j^\nu_\mu(q)$ is the Fourier transformed field of 
$j^\nu_{\mu{\bf r}}$ [Eq.~(\ref{jdef})]:
\begin{eqnarray}
j^\nu_0(q) &=& \frac{1}{\sqrt{\beta V}} \sum_k \phi^\dagger_k \sigma^\nu
  \phi_{k+q} ,  \nonumber \\ 
j^0_\mu(q) &=& \frac{1}{\sqrt{\beta V}} \sum_k \frac{1}{m} \biggl(k_\mu +
  \frac{q_\mu}{2} \biggr) \phi^\dagger_k \phi_{k+q} \,\,\, (\mu\neq 0) ,
\end{eqnarray}
where $V$ is the volume of the system and $V^{-1}\sum_{\bf k}={\int_{\bf k}}$
for 
$V\to\infty$. $k=({\bf k},i\omega)$ and $\sum_k = \sum_{{\bf k},\omega}$ where
$\omega$ is a fermionic Matsubara frequency. We have 
\begin{eqnarray}
{\Pi^{zz}_{00}(q)}  &=& - \frac{2}{\beta V} \sum_k [ G(k)G(k+q)-F(k)F(k+q)] ,
\nonumber \\
{\Pi^{z+}_{00}(q)}  &=& {\Pi^{z-}_{00}(-q)} = - \frac{2}{\beta V} \sum_k
G(k+q)F(k) , 
\nonumber \\ 
{\Pi^{++}_{00}(q)}  &=& {\Pi^{--}_{00}(q)} = - \frac{1}{\beta V} \sum_k
F(k)F(k+q) , 
\nonumber \\
{\Pi^{+-}_{00}(q)} &=&  \frac{1}{\beta V} \sum_k G(k)G(-k-q) ,
\nonumber \\ 
\Pi^{00}_{\mu\mu'}(q) &=& - \frac{2}{\beta V} \sum_k
\frac{1}{m^2} \biggl(k_\mu + \frac{q_\mu}{2} \biggr) \biggl(k_{\mu'} +
\frac{q_{\mu'}}{2} \biggr) \nonumber \\ && \times [G(k+q) G(k)+F(k+q) F(k)] ,
\label{pi1} 
\end{eqnarray} 
where $G$ and $F$ are the mean-field propagators [Eq.~(\ref{hfgf})]. The
correlation function $\langle j^0_\mu(q) j^\nu_0(-q) \rangle$ ($\mu\neq 
0$) vanishes. 

In the following, we consider the static limit $\Pi^{\nu\nu'}_{00}({\bf q}) = 
\Pi^{\nu\nu'}_{00}({\bf q},\omega_\nu=0)$. Performing the sum over Matsubara
frequency in (\ref{pi1}) in the $T=0$ limit, we obtain 
\begin{eqnarray}
{\Pi^{zz}_{00}({\bf q})} &=& {\int_{\bf k}} \frac{1}{{E_{\bf k}}+{E_{{\bf
	k}+{\bf q}}}} 
\biggl(1 - \frac{{\xi_{\bf k}}{\xi_{{\bf k}+{\bf q}}}}{{E_{\bf k}}{E_{{\bf
	k}+{\bf 
	q}}}} 
+ \frac{{\Delta_0^{\rm HS}}^2}{{E_{\bf k}}{E_{{\bf k}+{\bf q}}}} \biggr) , 
\nonumber \\  
{\Pi^{z+}_{00}({\bf q})}  &=& {\Pi^{z-}_{00}({\bf q})} = {\int_{\bf k}}
	\frac{\Delta_0^{\rm HS}{\xi_{\bf k}}}{({E_{\bf 
      k}}+{E_{{\bf k}+{\bf q}}}){E_{\bf k}}{E_{{\bf k}+{\bf q}}}} ,  
\nonumber \\
{\Pi^{++}_{00}({\bf q})} &=& {\Pi^{--}_{00}({\bf q})} = - {\int_{\bf k}}
	\frac{{\Delta_0^{\rm 
      HS}}^2}{2({E_{\bf k}}+{E_{{\bf k}+{\bf q}}}){E_{\bf k}}{E_{{\bf k}+{\bf
	q}}}} ,  
\nonumber \\
{\Pi^{+-}_{00}({\bf q})} &=& {\int_{\bf k}} \frac{1}{2({E_{\bf k}}+{E_{{\bf
	k}+{\bf q}}})} 
\biggl( 1 + 
\frac{{\xi_{\bf k}}{\xi_{{\bf k}+{\bf q}}}}{{E_{\bf k}}{E_{{\bf k}+{\bf q}}}}
\biggr) . 
\end{eqnarray} 
The correlation function $\Pi^{00}_{\mu\mu'}({\bf q})$ vanishes for ${\bf
q}=0$. Since $j^0_{\mu{\bf r}}$ multiplies $\partial_\mu\theta_{\bf r}$ in the
action $S'$, it is sufficient to consider $\Pi^{00}_{\mu\mu'}({\bf q}=0)$ to
obtain the effective action $S[\rho,\Delta]$ to order $(\partial_\mu
\theta_{\bf r})^2$.  

We next expand the correlations to order ${\cal O}(|{\bf q}|^2)$. Writing
${\xi_{{\bf k}+{\bf q}}}={\xi_{\bf k}} + {X_{{\bf k},{\bf q}}}$ with ${X_{{\bf
      k},{\bf q}}}={\bf k}\cdot{\bf q}/m+|{\bf 
  q}|^2/2m$, we 
obtain
\begin{eqnarray}
{\Pi^{zz}_{00}({\bf q})} &=& {\int_{\bf k}} \biggl[ \frac{{\Delta_0^{\rm
	HS}}^2}{{E^3_{\bf 
	k}}} + \biggl( 
  -\frac{3{\xi_{\bf k}}}{2{E^3_{\bf k}}} + \frac{3\xi^3_{\bf k}}{2E^5_{\bf k}}
	\biggr) 
  {X_{{\bf k},{\bf q}}} 
\nonumber \\ &&  
+ \biggl(-\frac{1}{2{E^3_{\bf k}}} + \frac{3{\xi^2_{\bf k}}}{E^5_{\bf k}} -
\frac{5\xi^4_{\bf k}}{2E^7_{\bf k}} \biggr) {X^2_{{\bf k},{\bf q}}} \biggr] ,
\nonumber \\ 
\Pi^{z\pm}_{00}({\bf q}) &=& {\int_{\bf k}} \frac{\Delta_0^{\rm
    HS}{\xi_{\bf k}}}{2E^3_{\bf k}} 
\biggl( 
  1 -\frac{3{\xi_{\bf k}}}{2{E^2_{\bf k}}} {X_{{\bf k},{\bf q}}} \nonumber \\
	&& 
- \frac{3{E^2_{\bf k}}-10{\xi^2_{\bf k}}}{4E^4_{\bf k}} {X^2_{{\bf k},{\bf
	q}}} \biggl) ,  
\nonumber \\ 
{\Pi^{++}_{00}({\bf q})} &=& {\Pi^{--}_{00}({\bf q})} \nonumber \\ 
&=& - {\int_{\bf k}} \frac{{\Delta_0^{\rm HS}}^2}{4{E^3_{\bf k}}} \biggl( 1 -
\frac{3{\xi_{\bf k}}}{2{E^2_{\bf k}}} {X_{{\bf k},{\bf q}}} \nonumber \\ &&
- \frac{3{E^2_{\bf k}}-10{\xi^2_{\bf k}}}{4{E^2_{\bf k}}} {X^2_{{\bf k},{\bf
	q}}} \biggr)  
\nonumber \\ 
{\Pi^{+-}_{00}({\bf q})} &=& \frac{1}{2} {\int_{\bf k}} \biggl[
	\frac{1}{2{E_{\bf k}}} + 
  \frac{{\xi^2_{\bf k}}}{2{E^3_{\bf k}}} + 
  \biggl( \frac{{\xi_{\bf k}}}{4{E^3_{\bf k}}} - \frac{3{\xi^3_{\bf
	k}}}{4E^5_{\bf k}} 
  \biggr) 
  {X_{{\bf k},{\bf q}}}    
\nonumber \\
&& + \biggl( - \frac{1}{8{E^3_{\bf k}}} - \frac{7{\xi^2_{\bf k}}}{8E^5_{\bf
    k}} + 
    \frac{5\xi^4_{\bf k}}{4E^7_{\bf k}} \biggr) {X^2_{{\bf k},{\bf q}}}
	\biggr] . 
\label{pi2}
\end{eqnarray} 

\subsection{Strong-coupling limit ${\rho_0 a^3}\ll 1$} 

In the strong-coupling limit, the chemical potential $\mu$ is negative. We
then have
\begin{eqnarray}
&& {\int_{\bf k}} \frac{1}{\xi_{\bf k}} = 
\frac{m\Lambda}{\pi^2} - \frac{m^{3/2}|\mu|^{1/2}}{\sqrt{2}\pi} , 
\nonumber \\ &&
\int_0^\infty dk \frac{k^4}{\xi^3_{\bf k}} = 
\frac{3\pi m^{5/2}}{2\sqrt{2}|\mu|^{1/2}} .
\label{B5}
\end{eqnarray}
Other useful relations are obtained by differentiating Eqs.~(\ref{B5})  with
respect to $\mu$. Note that $\Lambda$ is sent to infinity whenever the
integral over ${\bf k}$ converges. In the strong-coupling limit, the
small parameter expansion is ${\Delta_0^{\rm HS}}^2/|\mu|^2\sim
{\rho_0 a^3}$. Approximate 
expressions of the mean-field correlation functions can be obtained by
expanding Eqs.~(\ref{speq1}) and (\ref{pi2}) in power of ${\Delta_0^{\rm
HS}}^2$ and using Eqs.~(\ref{B5}) (as well as those obtained from (\ref{B5})
by differentiating 
with respect to $\mu$). A straightforward (but somewhat lengthly) calculation
then gives Eq.~(\ref{muhf}) and 
\begin{eqnarray}  
{\Pi^{zz}_{00}({\bf q})} &=& \rho_0 m a^2 \biggr(1-4\pi{\rho_0 a^3} -
\frac{{|{\bf q}|}^2 a^2}{4} 
\biggl) , 
\nonumber \\ 
\Pi^{z\pm}_{00}({\bf q}) &=& \biggl( \frac{\rho_0 m^2a}{4\pi} \biggr)^{1/2}
\biggl( 1 - \frac{7}{4} \pi{\rho_0 a^3} - \frac{{|{\bf q}|}^2 a^2}{12} \biggl)
, 
\nonumber \\
{\Pi^{++}_{00}({\bf q})} &=& - \frac{\rho_0 ma^2}{4} \biggl( 1 - 4\pi{\rho_0
  a^3} - 
\frac{5}{16}{|{\bf q}|}^2 
a^2 \biggr) , 
\nonumber \\
{\Pi^{+-}_{00}({\bf q})}  &=& \frac{1}{g} - \frac{\rho_0 ma^2}{4} ( 1 -
4\pi{\rho_0 a^3}) - \frac{ma{|{\bf q}|}^2}{32\pi} .
\end{eqnarray}


\begin{thebibliography}{99}
\bibitem{note2} In Ref.~\onlinecite{Depalo99}, it was (incorrectly) stated
  that the coefficient $1/32$ in the term $(\boldsymbol{\nabla}\rho_{\bf
  r})^2/32\rho_0 m$ is not 
  obtained from the expansion about the mean-field state, so that the
  action (\ref{action5}) is recovered only in the strict hydrodynamic limit
  where only the gradient term $\rho_{\bf r}({\boldsymbol{\nabla}\theta_{\bf
  r}})^2/8m$ contained in
  $|\boldsymbol{\nabla}{\psi_{\bf r}}|^2/4m$ is retained, while the term
  $(\boldsymbol{\nabla}\rho_{\bf r})^2/32m\rho_{\bf r}$ is ignored.
\end{thebibliography}
\end{document}